\documentclass[conference]{IEEEtran}
\ifCLASSINFOpdf
\else
\fi
\usepackage[pdftex]{graphicx}

\IEEEoverridecommandlockouts

\usepackage[overlay,absolute]{textpos}

\begin{document}

\begin{textblock*}{10in}(25mm, 10mm)
{\textbf{Ref:} \emph{International Joint Conference on Neural Networks (IJCNN)}, pages 1--8, Killarney, Ireland, July 2015.}
\end{textblock*}

\title{DeepSign: Deep Learning for Automatic Malware Signature Generation and Classification\textsuperscript{*}\thanks{\textsuperscript{*}The support of the KABARNIT Consortium under the Israeli Ministry of Industry, Trade and Labor is gratefully acknowledged.}}

\author{\IEEEauthorblockN{Eli (Omid) David\textsuperscript{$\dagger$} \thanks{\textsuperscript{$\dagger$}Also affiliated with the School of Computer Science, Tel Aviv University, Israel.}}%\IEEEauthorrefmark{1}}
\IEEEauthorblockA{Dept. of Computer Science\\
Bar-Ilan University\\
Ramat-Gan 52900, Israel\\
Email: mail@elidavid.com}
\and
\IEEEauthorblockN{Nathan S. Netanyahu\textsuperscript{$\ddagger$} \thanks{\textsuperscript{$\ddagger$}Also affiliated with the Gonda Brain Research Center at Bar-Ilan University, and the Center for Automation Research, University of Maryland at College Park, MD.}}
\IEEEauthorblockA{Dept. of Computer Science\\
Bar-Ilan University\\
Ramat-Gan 52900, Israel\\
Email: nathan@cs.biu.ac.il}
}

%Telephone: (800) 555--1212, Fax: (888) 555--1212}


\maketitle

\begin{abstract}
%\boldmath
This paper presents a novel deep learning based method for automatic malware signature generation and classification. The method uses a deep belief network (DBN), implemented with a deep stack of denoising autoencoders, generating an invariant compact representation of the malware behavior. While conventional signature and token based methods for malware detection do not detect a majority of new variants for existing malware, the results presented in this paper show that signatures generated by the DBN allow for an accurate classification of new malware variants. Using a dataset containing hundreds of variants for several major malware families, our method achieves 98.6\% classification accuracy using the signatures generated by the DBN. The presented method is completely agnostic to the type of malware behavior that is logged (e.g., API calls and their parameters, registry entries, websites and ports accessed, etc.), and can use any raw input from a sandbox to successfully train the deep neural network which is used to generate malware signatures.

\end{abstract}


\begin{IEEEkeywords}
Deep Learning, Deep Belief Network, Autoencoders, Malware, Automatic Signature Generation
\end{IEEEkeywords}

% \fi
\IEEEpeerreviewmaketitle

\section{Introduction}

Despite the nearly exponential growth in the number of new malware (e.g., Panda Security reports that on average 160,000 new malware programs appeared every day in 2013 \cite{lopez14}), the method for defending against these threats has largely remained unchanged. Anti-virus solutions detect the malware, analyze it, and generate a special handcrafted signature which is released as an update to their clients. This manual analysis phase typically takes a long time, during which the malware remains undetected and keeps infecting new computers. Additionally, even when detected, the authors of malware programs usually make some minimal changes to their code, so that the new variant is undetected by the anti-virus software. This ``cat and mouse'' game between malware developers and anti-virus companies goes on for many years for most major malware programs, and with each release of a new variant, thousands of computers are infected.

Several methods have been proposed for automatic malware signature generation, e.g., signatures based on specific vulnerabilities, payloads, honeypots, etc. A major problem associated with these methods is that they target specific aspects of the malware, thus allowing the malware developers to create a new undetected variant by modifying small parts of their software. For example, a malware spreading through the use of a specific vulnerability found in Windows operating system, can use another vulnerability in the system to spread, thus evading vulnerability-based signatures. 

In this paper we present a novel method for signature generation which does not rely on any specific aspect of the malware, thus being invariant to many modifications in the malware code (i.e., the proposed approach is capable of detecting most new variants of any malware). The method relies on training a \emph{deep belief network} (DBN) \cite{hinton06a}, i.e., a deep unsupervised neural network, which would create an invariant compact representation of the general \emph{behavior} of the malware. In recent years DBNs have proven successful in generating invariant representations for challenging domains, and our method attempts to use similar principles for generating invariant representations for malware.

The proposed method consists of the following steps in the unsupervised training phase: Given a dataset of malware programs, run each program in a sandbox to generate a text file containing the behavior of the program. Then, parse the sandbox text file and convert it to a binary bit-string to feed it to the neural network. Next, a deep belief network implemented using deep denoising autoencoders is trained by layer-wise training. The training is completely unsupervised, and the network is not aware of the labels of each sample. The DBN has eight layers, and its output layer contains 30 neurons. Thus, the resulting deep network basically generates a signature containing 30 floating point numbers for each program run in a sandbox.

We use a large dataset, containing several major malware categories and several hundred variants for each. The trained DBN generates a signature for each malware sample. The quality and representation power of these generated signatures is examined by running several supervised classification methods on them. The results show that a deep neural network achieves 98.6\% classification accuracy when tested on unseen data, which attests to the representation power of the signatures due to DBN.

In the next section we review several previous approaches for automatic signature generation. In Section III we describe our approach, and Section IV presents implementation details and experimental results. Section V contains our concluding remarks.

\section{Related Work}

It is very difficult to successfully generate signatures which can be used to prevent new attacks, and so the conventional methods are usually ineffective against zero-day malware \cite{filiol07, tang05, wang04}. Several approaches have been suggested to improve the signature generation process.  Here we briefly review  several of them.

Several methods which try to cope with new malware variants do so by analyzing the traffic (assuming that traffic patterns do not change substantially for each variant of the malware). \emph{Autograph} \cite{kim04} records source and destination of connections attempted from outside the network (inbound connections). An external source is considered to be a scanner if it has made more than a prespecified number of attempts to connect to an IP address in the network. After deeming this external source a scanner, and thus potentially malicious, Autograph selects the most frequent byte sequence from the network traffic of this source and uses it as its signature. A scanner malware already signed by Autograph can evade detection by modifying its most frequent byte sequence. A similar approach for signature generation based on network traffic is  \emph{Honeycomb} \cite{kreibich04}, which analyzes the traffic on the honeypot. Honeycomb uses largest common substrings (LCS) to generate signatures and measure similarities in packet payloads. The \emph{PAYL sensor} \cite{wang04} monitors the flow of information in the network and tries to detect malicious attacks using anomaly detection, assuming that the packets associated with zero-day attacks are distinct from normal network traffic. The \emph{Nemean architecture} \cite{yegneswaran05} is a semantic-aware Network Intrusion Detection System (NIDS) which normalizes packets from individual sessions in the network and renders semantic context. A signature generation component clusters similar sessions and generates signatures for each cluster. Another semantic-aware method is \emph{Amd} \cite{christodorescu09}, which generates semantic-aware code templates and specifies the conditions for a match between the templates and the programs being checked. \emph{Polygraph} \cite{newsome05} generates content based signatures that use several substring signatures (tokens), to expand the detection of malware variants. \emph{EarlyBird} \cite{singh04} sifts through the invariant portion of a worm's content that will appear frequently on the network as it spreads or attempts to spread. \emph{Netspy} \cite{wang04} also uses the invariant portion of network traffic generated by malware to generate a signature.

The majority of anti-virus programs reply on analyzing the executable file to determine whether it is a malware. As Filiol and Josse \cite{filiol07} establish, most current anti-virus programs do not detects variant of malware. They propose a method for automatic signature generation by analyzing the executable's code and substrings, and measure statistical distribution of code across variants of malware. Their experiments were performed on short (small sized) malware such as Nimda, Code Red/Code Red II, MS
Blaster, Sober, Netsky and Beagle. This method is less accurate when applied to larger malware. Most real world malware are large, containing many modules and sub-modules, and so a statistical analysis would not be sufficient to accurately classify them. \emph{Auto-Sign} \cite{tahan10} generates a list of signatures for a malware by splitting its executable to segments of equal sizes. For each segment a signature is generated, and the list of signatures is subsequently ranked. This method is more resilient to small modifications in the executable, but a malware can evade this method by encrypting the executable (which is a simple and popular method for many malware programs), and thus evading any method which inspects the executable file for signature comparison.

Since current approaches mostly rely on specific behavior of malware for signature generation (e.g., specific network traffic, or specific substrings in executable, etc.), new malware variants could be created with minimal modifications, such that they would not be detected by the conventional methods. In the next section we propose a method for signature generation based on the \emph{behavior} of the program, without focusing on any specific aspect of the executable or network traffic, thus making it difficult for a malware variant to evade detection.

\section{Proposed Signature Generation Method}

This section provides our novel approach for signature generation. The main question we are trying to answer is the following: \emph{Is it possible to generate a signature for a program that represents its behavior, and is invariant to small scale changes?} In recent years deep learning methods have proven very successful in accomplishing this very task in computer vision. Deep neural networks are trained to create invariant representations of objects, so that even when the object is in a different position, size, contrast, angle, etc., the network still detects the object correctly. These networks have achieved under 10\% error in the difficult task of ImageNet \cite{krizhevsky12,simonian14}. Unsupervised versions of these networks have been developed as well, e.g. \cite{lee08,le12}, where deep belief networks were training by merely exposing the networks to images randomly taken from YouTube videos. Krizhevsky and Hinton \cite{krizhevsky11} used deep autoencoders to create short binary codes for images based on their content (e.g., pictures containing elephants will have similar codes, etc.).

Our method uses these principles and applies them for modeling the behavior of programs (and specifically, malware). The goal is that the obtained representation would be invariant to small scale changes, and thus capable of detecting most variants of malware\footnote{Note that there are many similarities between our approach and that of  Krizhevsky and Hinton \cite{krizhevsky11}, as both use deep autoencoders to create short signatures for the \emph{content}; in our case the content is the high level behavior of the program (and not specific low level features such as strings in the executable), and in Krizhevsky and Hinton's case, it is the high level objects appearing in the image (and not low level features based on pixels in the image).}. To accomplish this goal, we first need to find a way to represent the behavior of a program as a fixed sized vector, which would be the input to the neural network. We will then train a deep belief network which would produce invariant representations of the input. The output of the DBN will be the signature for the malware.

\subsection{Program Behavior as Binary Vector}

\begin{figure}
	\centering
	\includegraphics[width=1\columnwidth]{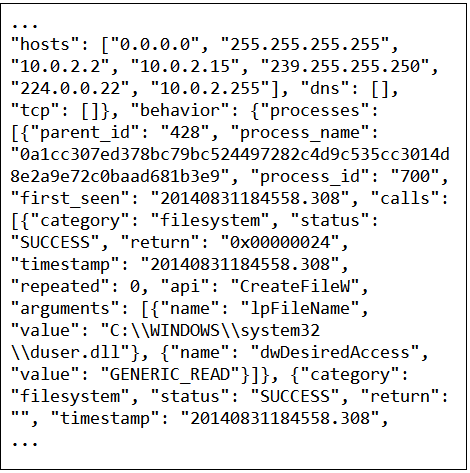}
	\caption{A snippet from the log file generated by Cuckoo sandbox.}
	\label{fig:cuckoo}
\end{figure}

Behavior of programs (and specifically malware) is typically recorded by running the programs in a sandbox. A sandbox is a special environment which allows for logging the behavior of programs (e.g., the API function calls, their parameters, files created or deleted, websites and ports accessed, etc.) The results are saved in a file (typically a text file). Figure \ref{fig:cuckoo} shows a snippet of logs recorded by a sandbox. Sandbox records are usually analyzed manually, trying to learn information that would assist in creating a signature for the malware (see Section II). 

The simplest method for converting the sandbox generated text file to a fixed size string is using one of the methods common in natural language processing (NLP). Of these methods, the simplest yet is unigram (1-gram) extraction. For example, given a dataset of text samples, find the 5,000 most frequent words in the text (these words would comprise the dictionary), and then for each text sample check which of these 5,000 words are present. Thus, each text sample is represented as a 5,000 sized bit-string. Unlike language text files, sandbox files contain a variety of information, and require several preprocessing stages to extract the useful content (e.g., string after \texttt{"api"} tag contains the name of function call, etc.). However, in order to remain as domain agnostic as possible, we propose to treat the sandbox file as a simple text file, and extract unigrams without any preprocessing. That is, all the markup and tagged part of the files are extracted as well (e.g., given \texttt{"api": "CreateFileW"}, the terms extracted are \texttt{"api":} and  \texttt{"CreateFileW"}, completely ignoring what each part means). While this may sounds absurd (intentionally adding useless noise where it can be easily removed), this should not pose a problem, since the learning system (described below) should easily learn to ignore these irrelevant parts. Specifically, our method follows the following simple steps to convert sandbox files to fixed size inputs to the neural network: (1) For each sandbox file in the dataset, extract all unigrams, (2) remove the unigrams which appear in all files (contain no information), (3) for each unigram count the number of files in which it appears, (4) select top 20,000 with highest frequency, and (5) convert each sandbox file to a 20,000 sized bit string, by checking whether each of the 20,000 unigrams appeared in it. In other words, we first define which words (unigrams) participate in our dictionary (analogous to the dictionaries used in NLP, which usually consist of the most frequent words in a language), and then for each sample we check it against the dictionary for the presence of each word and thus produce a binary vector.

\subsection{Training a Deep Belief Network}

The previous subsection described a simple method for converting the behavior of a computer program to a fixed size binary vector. As we discussed previously, most malware variants make small changes in their code (i.e., small changes in behavior), which is sufficient to evade the classical signature generation methods. We would like to generate a signature for each program which is resilient to these small changes (an invariant representation, similar to those used for computer vision). In order to achieve this goal, we create a deep belief network (DBN) by training a deep stack of denoising autoencoders.

An autoencoder is an unsupervised neural network which sets the target values (of the output layer) to be equal to the inputs, i.e., the number of neurons at the input and output layers is equal, and the optimization goal for output neuron $i$ is set to equal $\mathbf{x}_i$, which is the value of the input neuron
$i$. A hidden layer of neurons is used between the input and output layers, and the number of neurons in the hidden layer is usually set to fewer than those in the input and output layers, thus creating a bottleneck, with the intention of forcing the network to learn a higher level representation of the input. That is, for each input $\mathbf{x}$, it is first mapped to a hidden layer $\mathbf{y}$, and the output layer tried to reconstruct $\mathbf{x}$. The weights of the encoder layer ($\mathbf{W}$) and the weights of the decoder layer ($\mathbf{W'}$) can be tied (i.e., defining $\mathbf{W' = W}^T$). Autoencoders are typically trained using backpropagation with stochastic gradient descent \cite{rumelhart86,werbos74}.

\begin{figure}
	\centering
	\includegraphics[width=0.6\columnwidth]{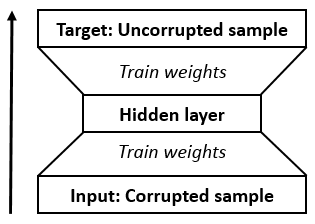}
	\caption{One layer of denoising autoencoder during training.}
	\label{fig:denoising}
\end{figure}

Recently it has been demonstrated that \emph{denoising autoencoders} \cite{vincent10} generalize much better than basic autoencoders in many tasks. In denoising autoencoders each time a sample is given to the network, a small portion (usually a ratio of about 0.1 to 0.2) of it is corrupted by adding noise (or more often by zeroing the values). That is, given an input $\mathbf{x}$, first it is corrupted to $\mathbf{\tilde{x}}$ and then given to the input layer of the network. The objective function of the network in the output layer remains generating $\mathbf{x}$, i.e., the uncorrupted version of the input (see Figure \ref{fig:denoising}). This approach usually works better than basic autoencoders due to diminishing the overfitting in the network. By having to recreate the uncorrupted version of the input, the network is forced to generalize better, and determine more high level patterns. Additionally, since the network rarely receives the same input pattern more than once (each time sees a corrupted version only), there is a diminished risk of overfitting (though it still takes place). Finally, using denoising autoencoders the hidden layer need not necessarily be smaller than the input layer (in basic autoencoder such a larger hidden layer may result in simply learning the identity function). Note that the noise is added only during training. In prediction time the network is given the uncorrupted input (i.e., similar to basic autoencoder).

When an autoencoder's training is complete, we can discard the decoder layer, fix the values of the encoder layer (so the layer can no longer be modified), and treat the output of the hidden layer as the input to a new autoencoder added on top of the previous autoencoder. This new autoencoder can be trained similarly. Using such layer-wise unsupervised training, deep stacks of autoencoders can be assembled to create deep neural networks consisting of several hidden layers (forming a deep belief network). Given an input, it will be passed through this deep network, resulting in high level outputs. In a typical implementation, the outputs may then be used for supervised classification if required, serving as a compact higher level representation of the data.

\begin{figure*}
\centering
\includegraphics[width=1.0\textwidth]{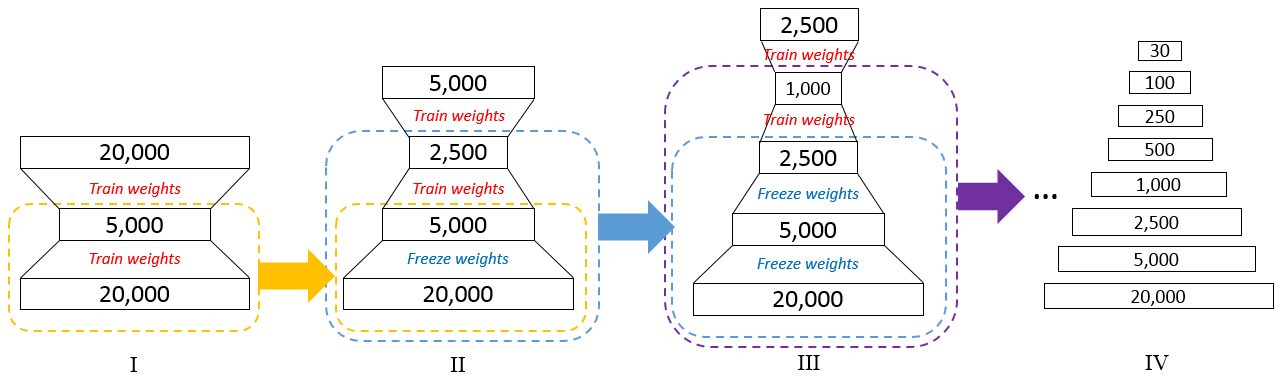}
\caption{Illustration of DBN training. (I) Train the first autoencoder layer. (II) Use the weights from the previous layer, and build a new autoencoder on top of them. The weights taken from the first autoencoder are frozen. (III) Similarly, take the weights from the previous layer and add a new layer. (IV) Proceed with layer-wise training until training all the eight layers for the final DBN.}
\label{fig:dbn}
\end{figure*}

In our approach we train a deep denoising autoencoder consisting of eight layer: 20,000--5,000--2,500--1,000--500--250--100--30. At each step only one layer is trained, then the weights are ``frozen'', and the subsequent layer is trained, etc. (see Figure \ref{fig:dbn}). At the end of this training phase, we have a deep network which is capable of converting the 20,000 input vector into 30 floating point values. We regard these 30-sized vector as the ``signature'' of the program. Note that the network is trained only using the samples in the training set, and for all future samples it will be run in prediction mode, i.e., receiving the 20,000-sized vector it will produce 30 output values, without modifying the weights.

The next section provides implementation details and experimental results, and demonstrates that the resulting 30-sized vector (i.e., the signature) indeed provides a good invariant representation of the malware.

\section{Implementation and Experimental Results}

In this section we first describe the malware dataset and the properties of the sandbox that was used, then provide the details of the trained neural network, and finally present our experimental results.

\subsection{Malware Dataset and Sandbox}

Our dataset consists of six major categories of malware, and 300 variants in each category, for a total of 1,800 samples. Each of these six malware categories spread massively worldwide and caused a tremendous damage. Hundreds of variants of them were created, each time modifying some parts of the malware to evade anti-virus programs. These new variants remained undetected until they were manually detected, analyzed, and a signature was generated for them. 

The six malware categories used are Zeus, Carberp, SpyEye, Cidox, Andromeda, and DarkComet. All of these six malware families are used to carry out a wide range of criminal tasks, and have infected millions of computers worldwide. Several crackdowns by the FBI and other law enforcement agencies in numerous countries have resulted in the arrest of more than a hundred persons involved with development and use of these malware, but their variants are widely used to the present day. The following is a brief description of the six malware classes.

\textbf{Zeus.} Probably the most widely used Trojan for cyber crime (especially for stealing financial information). It was first detected in July 2007 and is still widely used both in its original format and in thousands of variants which are continuously introduced to evade anti-viruses. It is estimated that in the US alone it has infected about 3.6 million computers. Several of the top malware programs used for stealing banking information are based on variations of Zeus. The entire source code of Zeus is freely available online, facilitating the creation of new malware based on it. 

\textbf{Carberp.} A widespread malware that silently downloads and installs other malware components to the infected system. It was first discovered in 2010 and reported as the sixth most popular malware for stealing financial information. Currently there are no clear estimates of the number of infected systems or amount of money earned by the developers, as this malware remains mostly underground. In its later versions it heavily incorporates Zeus code.

\textbf{SpyEye.} First reported in 2009 as a banking Trojan but it has been used to carry payloads for industrial espionage as well. It has infected 1.4 million computers worldwide since 2009, and the developers of this malware have made more than \$3.2 million in a six-month period alone. It specializes in stealing valuable personal information from the victim's computer, including banking login and passwords, credit card numbers, social security numbers, etc.

\textbf{Cidox.} A remote administrative tool (RAT) which is mainly used to control infected systems. This Trojan is not self-replicating, but is rather spread via manual targeting of victims. It is one of the first malware not hiding in the master boot record of Windows operating system, and instead, hides in network file system locations. It reconfigures the NTFS file system's program loader, thus becoming invisible in the file system.

\textbf{Andromeda.} One of the most widespread non-replicating spam bots, which mostly spreads via email-based infections. It was inactive for a certain duration, but has recently resurfaced with more sophisticated features. This malware was first identified in February 2007, and it was reported that most of the infected systems were in the European countries. Andromeda is highly modular, and can incorporate various modules (keylogger, screen capture, etc.).

\textbf{DarkComet.} A remote administration tool, first discovered in February 2012. It is used in a wide range of targeted attacks, and has the ability to take pictures via webcam, record conversations via a microphone attached to the PC, and gain full control of the infected machine. It is freely available online, and as a result, one of the most popular remote administration tools.

In 2011 the source code of Zeus was leaked, and since then many other malware have started incorporating its code into their program. As a result, at times it is difficult to categorize a new variant as either Zeus or one of the other malware families (e.g., variant of Carberp which uses many parts of Zeus code is commonly referred to as ``Zberp''). In this work we use the categories provided by Kaspersky anti-virus as our ground truth. That is, if Kaspersky classifies a malware as a variant of Carberp, then for our purposes that is the correct label (hence, the prediction task for our learning module is difficult, because the six different classes of malware are not completely separated.)

Each of the 1,800 programs in our dataset is run in Cuckoo sandbox\footnote{Available at \texttt{http://www.cuckoosandbox.org}}, the most popular open source sandbox tool for malware analysis. Cuckoo sandbox records native functions and Windows API call traces, details of files created and deleted from the filesystem, IP addresses, URLs and ports accessed by the program, registry keys written, etc. The result is saved in a text file in JSON file format (though note that as described in the previous section our approach is agnostic to the format of this text file, and completely ignores the formatting). Using the procedure described in the previous section each of these sandbox files is converted to a 20,000 sized bit-string, which is a rough fixed size representation of the raw sandbox text file.

Having converted all our dataset to 1,800 vectors (each of size 20,000), we randomly split them to 1,200 samples for training (200 samples from each of the six categories) and 600 samples for testing (100 samples from each category).

\subsection{Training the DBN}

As described in the previous section, we train a deep denoising autoencoder consisting of eight layers (20,000--5,000--2,500--1,000--500--250--100--30), with layer-wise training. To further regularize the network and prevent overfitting, we use \emph{dropout} \cite{hinton12}. Each time a new input is given to the network, each hidden unit is randomly omitted from the network with a probability of 0.5 (i.e., about half of the units in the hidden layer are omitted). The idea is that a hidden unit cannot rely on other hidden units being present. Dropout is essentially an efficient way for performing model averaging. Instead of training many separate networks and then applying each of these networks to the test data and calculating the average over the predictions (which is computationally expensive), random dropout makes it possible to train a huge number of different networks in a reasonable time. In prediction time, all the neurons in hidden layer are present, but their output is multiplied by 0.5 (halved). Note that in our case, training autoencoders, at each learning step we have only one hidden layer. For example, when training the first layer 20,000 to 5,000 to 20,000, then only the neurons in the hidden layer of 5,000 neurons are affected by dropout. During prediction, the output of each of these hidden units is halved.

Instead of using the standard logistic or tanh activation functions, we use \emph{rectified linear units} (ReLU) for the non-linearity function \cite{glorot11}.

\begin{math}
	f(x) = \max(0,x)
\end{math}

ReLU is widely used when training deep neural networks, usually resulting in faster convergence and diminishes the gradient vanishing problem, which especially affects deep networks \cite{bengio09}. 

Other parameters we use are: noise ratio of 0.2 for denoising autoencoders, 1000 training epochs (for each autoencoder layer), learning rate which starts at 0.001 and linearly decays to 0.000001, batch size of 20, and no momentum. We use an L2 penalty for network regularization. Note that each layer has an additional bias unit, which is connected to all the units in the subsequent layer.

Due to the large network size (e.g., only the layer connecting 20,000 input neurons to 5,000 neurons contains more than 100,000,000 weights which should be learned), we ran the network on an Nvidia GeForce GTX 680 graphics card (GPU). This reduced the training time to under two days.

\begin{figure}
	\centering
	\includegraphics[width=1.0\columnwidth]{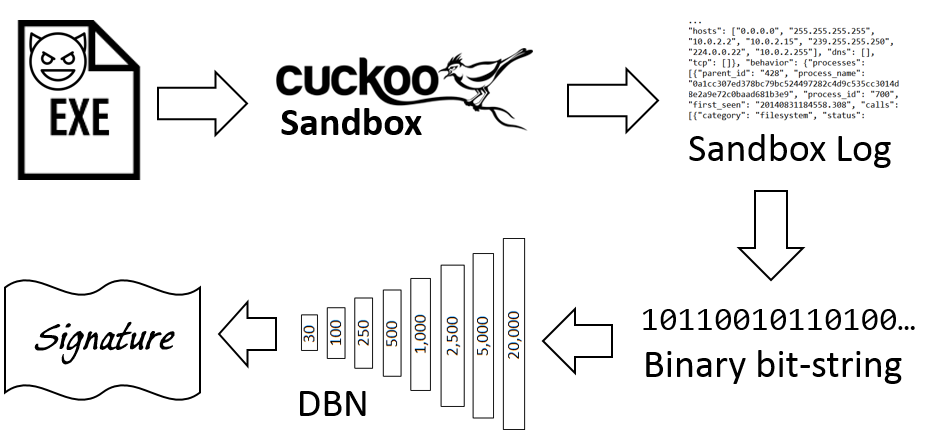}
	\caption{Illustration of all the stages from initial malware run in Sandbox to signature derivation using DBN.}
	\label{fig:process}
\end{figure}

Putting the above steps together, we have constructed an end-to-end method for automatic signature generation: The program is run in a sandbox, the sandbox file is converted to a binary bit-string which is fed to the neural network, and the deep neural network produces a 30-sized vector at its output layer, which we treat as the signature of the program. See Figure \ref{fig:process}.

\subsection{Experimental Results}

We now examine the quality of the generated signatures due to DeepSign. To do so, we feed all of our 1,800 vectors of size 20,000 to the DBN, and convert them to 30-sized representations (signatures). 

\begin{figure}
	\centering
	\includegraphics[width=1\columnwidth]{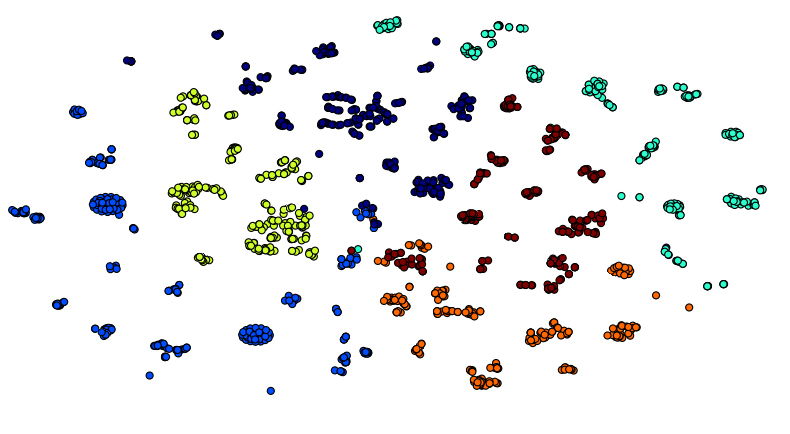}
	\caption{A 2-dimensional visualization of the malware signatures (each node is one malware signature), generated by the t-SNE dimensionality reduction algorithm. Each color corresponds to one of six malware categories. Note that the labels are used for coloring the nodes only, and otherwise the visualization is due to completely unsupervised DBN.}
	\label{fig:tsne}
\end{figure}

Figure \ref{fig:tsne} provides a two dimensional visualization of the data, where each node is one malware signature. The visualization is generated using the \emph{t-distributed stochastic neighbor embedding} (t-SNE) algorithm \cite{maaten08}, in this case reducing the dimensionality of the data from 30 (signature length) to 2. The goal of t-SNE is to reduce the dimensionality such that the closer two nodes are to each other in the original high dimensional space, the closer they would be in the 2-dimensional space. Note that the labels are used for coloring the nodes only, and otherwise the visualization is due to unsupervised DBN. The figure illustrates that variants of the same malware family are mostly clustered together in the signature space, demonstrating that the signatures due to DBN indeed capture invariant representations of malware. Some clustering errors are expected here (as can be seen in the visualization), since as explained in Subsection IV-A, many of these malware classes use parts of code from each other, and the distinction even amongst anti-virus detections is blurred. Here we use the labels given by Kaspersky anti-virus as the ground truth against which we measure the performance of our method.

To further measure the quality of this compact representation, we train a supervised classifier on the 30-sized vectors as follows: Train the classifier on the 1,200 vectors of size 30, and then predict on the 600 test vectors (of size 30). The higher the prediction accuracy is, the better the generated signatures are.

We first train an SVM classifier\footnote{We use the popular LIBSVM library \cite{chang11}.} using 1,200 signatures, and then use it to predict the correct labels (out of 6 possible) on the 600 prediction signatures. The resulting accuracy is 96.4\%. Alternatively, running a basic $k$-nearest neighbor algorithm (with $k=1$) where each of the 600 prediction samples are given the label of their nearest neighbor (Euclidean distance) from the 1,200 training samples, results in an accuracy of 95.3\%. This high accuracy obtained when training and predicting solely on the compact signature space attests to the fact that DeepSign generates meaningful signatures for the malware, resulting in successful detection of a high percentage of the malware variants generated with the purpose of evading classical anti-virus signatures.

Finally, to examine whether the classification accuracy in the supervised learning context can be improved, we use the weights of the trained neural network due to the DBN as the initial weights for a deep supervised neural network. The supervised network has exactly the same layers as the DBN, but with the addition of six neurons in the output layer (corresponding to six categories of malware). The neurons in this added output layer are softmax units, minimizing the cross-entropy loss function. Training this network on the 1,200 input training samples (using input noise = 0.2, dropout = 0.5, and learning rate = 0.001), and predicting on 600 test samples results in 98.6\% accuracy on test data, a relatively substantial improvement over the SVM results.

\section{Concluding Remarks}

In this paper we reviewed past approaches for generating signatures for malware programs, and proposed a novel method based on deep belief networks. Current approaches for malware signature generation use specific aspects of malware (e.g., certain network traffic normality or a substring in the program); thus, new malware variants easily evade detection by modifying small parts of their code.

Our proposed approach is inspired by the recent success in training deep neural networks which produce invariant representations. We first run the malware in a sandbox and then convert the sandbox log file to a long binary bit-string. This bit-string is fed to a deep 8-layered neural network which produces 30 values in its output layer. These values are used as the signature of the program. The experimental results show that the signatures produced by the DBN are highly successful for malware detection. These signatures can either be used in a completely unsupervised framework or used for supervised malware classification.

The results presented here demonstrate that unsupervised deep learning is a powerful method for generating high level invariant representations in domains beyond computer vision, language processing, or speech recognition; and can be applied successfully to challenging domains such as malware signature generation.

\end{document}